\documentclass[%
 aip,
 amsmath,amssymb,
 reprint,%
 citeautoscript,
]{revtex4-2}

\usepackage{amsbsy}
\usepackage{graphicx}
\usepackage{dcolumn}
\usepackage{bm}
\usepackage{multirow}
\usepackage[dvipsnames]{xcolor}
\usepackage{chemformula}
\usepackage{gensymb}
\usepackage{amsmath}
\usepackage[utf8]{inputenc}
\usepackage[T1]{fontenc}
\usepackage{mathptmx}
\usepackage{siunitx}
\usepackage{cancel}

\newlength{\CapLen}
\AtBeginDocument{\settoheight{\CapLen}{K}}

\newcommand\footnoteref[1]{\protected@xdef\@thefnmark{\ref{#1}}\@footnotemark}
\newcommand\ff{\boldsymbol{f}}

\begin{document}
\title{Molecular-orbital-based Machine Learning for Open-shell and Multi-reference Systems with Kernel Addition Gaussian Process Regression}

\author{Lixue Cheng}
\affiliation{Division of Chemistry and Chemical Engineering, California Institute of Technology, Pasadena, CA 91125, USA}
\author{Jiace Sun}
\affiliation{Division of Chemistry and Chemical Engineering, California Institute of Technology, Pasadena, CA 91125, USA}
\author{J. Emiliano Deustua}
\affiliation{Division of Chemistry and Chemical Engineering, California Institute of Technology, Pasadena, CA 91125, USA}
\author{Vignesh C. Bhethanabotla}
\affiliation{Division of Chemistry and Chemical Engineering, California Institute of Technology, Pasadena, CA 91125, USA}
\author{Thomas F. Miller III}
\email{tfm@caltech.edu.}
\affiliation{%
Division of Chemistry and Chemical Engineering, California Institute of Technology, Pasadena, CA 91125, USA
}%

\date{\today}
\begin{abstract}
We introduce a novel machine learning strategy, kernel addition Gaussian process regression (KA-GPR), in molecular-orbital-based machine learning (MOB-ML) to learn the total correlation energies of general electronic structure theories for closed- and open-shell systems by introducing a machine learning strategy. 
The learning efficiency of MOB-ML(KA-GPR) is the same as the original MOB-ML method for the smallest criegee molecule, which is a closed-shell molecule with multi-reference characters. 
In addition, the prediction accuracies of different small free radicals could reach the chemical accuracy of 1 kcal/mol by training on one example structure. Accurate potential energy surfaces for the \ch{H10} chain (closed-shell) and water OH bond dissociation (open-shell) could also be generated by MOB-ML(KA-GPR). 
To explore the breadth of chemical systems that KA-GPR can describe, we further apply MOB-ML to accurately predict the large benchmark datasets for closed- (QM9, QM7b-T, GDB-13-T) and open-shell (QMSpin) molecules.
\end{abstract}

\maketitle

\section{Introduction}
In recent years, a variety of machine learning (ML) approaches have emerged which promise to mitigate the cost of highly accurate electronic structure methods while preserving accuracy. \cite{Bartok2010,rupp2012fast,VonLilienfeld2013,hansen2013assessment,Ceriotti2014,ramakrishnan2015big,Tuckerman,kearnes2016molecular,Paesani2016, Behler2016,schutt2017quantum,schutt2017schnet,Smith2017,Welborn2018,wu2018moleculenet,Nguyen2018,Yao2018, Fujikake2018,unke2019physnet,Cheng2019,cheng2019regression,dick2020machine,deephf,qiao2020orbnet,qiao2020multi,hermann2020deep,deringer2021origins,christensen2021orbnet,husch2020improved,lee2020analytical,sun2021molecular,Karandashev2022,veit2020predicting,orbnetequi,klicpera2020fast,liu2022spherical,painn,faber2018alchemical,huang2020quantum,bartok2017machine,Welborn2018,Cheng2019,cheng2019regression,husch2020improved,lee2020analytical,sun2021molecular,cheng2022accurate,sun2022molecular}
The majority of these ML approaches are restricted to weakly-correlated, closed-shell organic molecules, and important recent advancements have not yet been carried over into the open-shell or strongly-correlated systems. 
In particular, only a few practical demonstrations of open-shell ML methods are developed, including the NN framework developed by Kulik and co-workers\cite{kulik2020making,duan2021putting,janet2019designing} and AIMNet-ME \cite{zubatiuk2021development,zubatyuk2021teaching}, to predict the energies of open-shell molecular systems at DFT accuracy. Although density functional theory is the most commonly used computational tool to compute the energies of these challenging systems, DFT methods are not able to systematically deliver results with controlled accuracy. \cite{grafenstein2002can,goodpaster2012density,johnson2017communication} Currently,  only a few ML approaches are able to provide accurate solutions to Schr\"{o}dinger equation beyond DFT levels for open-shell and strongly-correlated system, i.e., PauliNet\cite{hermann2020deep}, SpookyNet \cite{unke2021spookynet}, and DeepErwin \cite{scherbela2022solving}.

Molecular-orbital-based machine learning (MOB-ML) is a ML approach that provides state-of-the-art accuracies of post-Hartree-Fock (post-HF) theories at HF costs in various closed-shell benchmark systems\cite{Welborn2018,Cheng2019,cheng2019regression,husch2020improved,sun2021molecular,cheng2022accurate,sun2022molecular,lu2022fast}.
In this communication, we extend the current MOB-ML framework to the more complicated open-shell and strongly-correlated systems with multi-reference characters for general electronic structure theory using multi-reference configuration interaction (MRCI) as an example. By the kernel addition Gaussian process regression (KA-GPR), the MOB-ML models could be constructed by directly predicting the total correlation energies. The performance of MOB-ML is demonstrated on various chemical systems, including criegee, \ch{H10} chain, 9 small radicals, QM9\cite{ramakrishnan2014quantum,dipole_data}, QM7b-T, GDB-13-T \cite{Cheng2019,dataset,caltech_data}, water bond dissociation, and QMSpin datasets \cite{schwilk2020large,qmspin_data}.

\section{Theory and method}
\subsection{MOB-ML formalism for general electronic structure theory at ground state}

MOB-ML is a ML approach that predicts correlation energies using MOB features generated from HF computations  \cite{Welborn2018,Cheng2019,cheng2019regression,husch2020improved,sun2021molecular,cheng2022accurate,sun2022molecular}. Throughout this paper, we use capital letters $I,J,A,B\dots$ to represent spin orbitals, and lower-case letters $i,j,a,b\dots$ to represent spatial orbitals. Utilizing Nesbet's theorem, the correlation energy can be written exactly as a sum of contributions from occupied pairs of spin orbitals $(I,J)$, \cite{Nesbet1958} 
\begin{equation}
     E_\text{corr} = \sum_{IJ \in\text{occ}} \epsilon_{IJ}.
\end{equation}
For closed-shell systems, the pair energies can also be defined for each pair of occupied spatial orbitals $(i, j)$ as
\begin{equation}
    \epsilon_{ij}=\epsilon_{i\uparrow j\uparrow} + \epsilon_{i\uparrow j\downarrow} + \epsilon_{i\downarrow j\uparrow} + \epsilon_{i\downarrow j\downarrow}
\end{equation}
In previous studies \cite{Welborn2018,Cheng2019,cheng2019regression,husch2020improved,sun2021molecular,cheng2022accurate,sun2022molecular}, we have examined the special case of this general equation for closed-shell systems and performed all the calculations in terms of spatial orbitals.
However, MOB-ML can be straightforwardly applied to both closed-shell and open-shell systems with or without multi-reference characters at ground state by extending the spatial orbital formulation to the general form in spin orbitals.
Specifically, we seek to construct the map  
\begin{equation}
    \epsilon_{IJ} \approx \epsilon^\text{ML}(\ff_{IJ}),
\end{equation}
where $\ff_{IJ}$ denotes the MOB feature vector representing the spin orbitals $I$ and $J$ .
For general spin orbitals, the pair correlation energy is defined as 
\begin{equation} \label{eq:pair_energy}
    \epsilon_{IJ}=\sum_{AB \in \text{vir}} \langle \Phi_0 | \hat{H} | \Phi_{IJ}^{AB} \rangle \langle \Phi_{IJ}^{AB} | \Psi_0 \rangle,
\end{equation}
where $|\Phi_0\rangle$ is the HF ground state, $|\Phi_{IJ}^{AB}\rangle$ is the HF excited state by exciting orbitals $I, J$ to $A, B$, and $|\Psi_0\rangle$ is the true ground state.
According to Eq.~\ref{eq:pair_energy}, the pair energies can be calculated for any electronic structure method that has an explicit expression of the group state wavefunction and also satisfies $E=\langle \Phi_0 | \hat{H} | \Psi_0 \rangle$.
However, for methods not using HF orbitals, the pair energies are computationally expensive and also complicated to implement. For example, MRCI uses the multiconfigurational self-consistent field (MCSCF) orbitals. The calculation of each pair energy for any single Slater determinant would require the calculation of the overlap between two Slater determinants with different orbitals, which becomes 
\begin{equation}
    \langle a_1 a_2 \dots a_N | b_1 b_2 \dots b_N\rangle=\text{det}(S_{ab}),
\end{equation}
where
$\{a_i\}$, $\{b_i\}$ are the two sets of spatial orbitals, and $S_{ab}$ is their overlap matrix.
The computational cost to naively calculate all the pair energies scales as $O(N_{\text{orbitals}}^5 N_{\text{determinants}})$, which is extremely expensive due to the large amount of the Slater determinants in MRCI. 

\subsection{MOB-ML features generated from ROHF for open-shell systems}
\label{sec:openshell}
We use ROHF as the open-shell HF reference in MOB-ML, where the spin-up and down electrons share the same orbitals but have different occupations.
According to Ref. \onlinecite{husch2020improved}, the MOB-ML closed-shell features can be written as functions of Fock, Coulomb, and exchange matrices with number of occupation $n_\text{occ}$:
\begin{equation} \label{eq:features_closeshell}
    \ff = \ff(F, J, K; n_\text{occ}),
\end{equation}
where $n_{\text{occ}}$ is used to determine the split of occupied and virtual spaces.
We generalize the closed-shell MOB-ML features to open-shell systems by defining the spin-up features and spin-down features as
\begin{equation} \label{eq:features_openshell}
\begin{aligned}
    \ff^{\uparrow} &= \ff(F, J, K; n_\text{occ}^{\uparrow})
    \\
    \ff^{\downarrow} &= \ff(F, J, K; n_\text{occ}^{\downarrow}).
\end{aligned}
\end{equation}
Although the $F, J, K$ matrices for spin up and down electrons are the same in ROHF, the different occupation numbers result in different feature elements for the two spin types.

\subsection{Kernel addition Gaussian process regression (KA-GPR) for closed-shell and open-shell systems}

Since the definition of spin-up and spin-down features in Eq. \ref{eq:features_openshell} reduce to Eq. \ref{eq:features_closeshell} for closed-shell molecules, we present the formalism of KA-GPR for open-shell systems and treat closed-shell systems as special cases.
Let $\ff_{pi}^s$ be the MOB diagonal feature vector of the spatial orbital $\phi_i$, spin $s$ in molecule $M_p$.
KA-GPR regresses energies $E_p$ as a function of its feature vectors $\{\ff_{pi}^s | i\in \text{occupied orbitals}, s  = \uparrow, \downarrow\}$.
Given a basic kernel $K: \ff_{pi}^s, \ff_{qj}^s \rightarrow K(\ff_{pi}^s, \ff_{qj}^s)$, the kernel matrix element of two input molecules $M_p$ and $M_q$ in KA-GPR is defined as
\begin{equation}
     K^{\text{add}}_{pq} = \frac{1}{2} \sum_{ijs} K(\ff_{pi}^s, \ff_{qj}^s ).\label{eq:kernel_open}
\end{equation}
For closed-shell system, it reduces to 
\begin{equation}
     K^{\text{add}}_{pq} = \sum_{ij} K(\ff_{pi}, \ff_{qj} ).\label{eq:kernel_close}
\end{equation}

Given the training molecules $\{M_p\}$,
the predicted correlation energy for a test molecule $p^\prime$ is
\begin{equation}
    E_{p^\prime}^{\text{corr,ML}} = \sum_{p} K^\text{add}_{p^\prime p} \omega_p,
\end{equation}
where the $\omega$ vector is solved by
\begin{equation}
    \sum_{q} (K^\text{add}_{pq} + \sigma_n^2 \delta_{pq}) \omega_q = E_p^{\text{corr}},
\end{equation}
where $\sigma_n^2$ is the Gaussian noise.
KA-GPR implies an exact decomposition of the predicted correlation energy $E_{p^\prime}^{\text{corr,ML}}$ to spin orbitals (represented by $(i, s)$) given by
\begin{equation}
    E_{p^\prime}^{\text{corr,ML}} = \sum_{is} \left(\sum_{pj} K(\ff_{p^\prime i}^s, \ff_{pj}^s) \omega_p\right).
\end{equation}

In the MOB-ML (KA-GPR) framework, the kernel sum only includes the diagonal MOB features and omits the contributions from the off-diagonal part because the pair energies obtained from the off-diagonal part could be directly folded into the diagonal part:
\begin{equation}
    \epsilon_I = \sum_J \epsilon_{IJ}
\end{equation}
This framework could be considered as an implicit one-body decomposition of the correlation energies. The successes of OrbNet \cite{qiao2020orbnet,orbnetequi} and QML (MO, $\Delta$-learning)\cite{Karandashev2022} approaches, which both use one-body features, also exhibit that it is sufficient for ML to provide accurate predictions with implicit one-body decompositions. 
Also, the kernel construction costs of including off-diagonal contributions will significantly increase the computational cost of the kernel construction from $O(N^2)$ to $O(N^4)$, where $N$ is the number of occupied orbitals.

\begin{figure}[htbp]
    \centering
    \includegraphics[width=0.9\columnwidth]{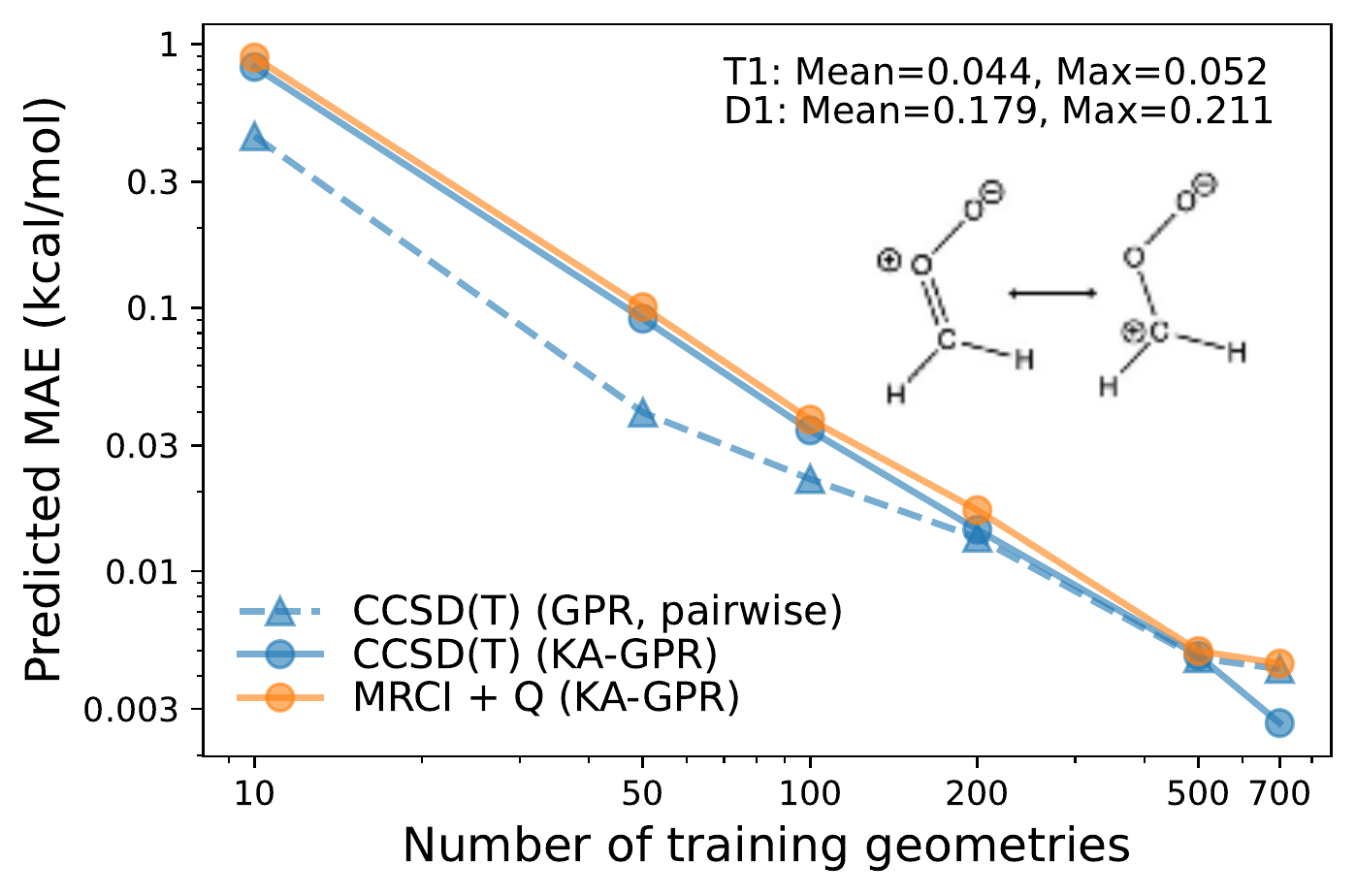}
    \caption{Mean absolute errors (MAEs) of predicted correlation energies for 100 test criegee geometries using MOB-ML. 
    For the reference theory of CCSD(T)/cc-pVTZ (blue), we display the results from learning the pair energies by GPR and total correlation energies by KA-GPR, respectively. For the reference theory of MRCI+Q/cc-pVTZ (orange), only the results learnt by KA-GPR using total correlation energy labels are reported. All the features are generated from the computation results of RHF/cc-pVTZ. 
    All the MAEs are plotted as functions of the number of training geometries on a logarithm scale ("learning curves"). The Lewis structures of the criegee molecule are also shown. All the prediction MAEs are also listed in the Supporting Information Table S1.}    
    \label{fig:criegee}
\end{figure}

\section{Computational details}
\subsection{Dataset preparation}
The criegee dataset contains 800 thermalized structures for the smallest criegee molecule (\ch{CH2O}). The small radical dataset includes 9 open-shell radicals and each radical contains 200 thermalized structures.
All the thermalized structures above are sampled at 50 fs intervals from individual \textit{ab initio} molecular dynamics (AIMD) trajectories at 350 K performed with the \textsc{Q-Chem} 5.0 package\cite{Shao2015}
using the B3LYP\cite{Vosko1980,Lee1988,Becke1993,Stephens1994}/6-31G*\cite{Hariharan1973} level of theory. For the water bond dissociation dataset, 50 thermalized structures are first randomly selected from the MOB-ML water dataset \cite{Cheng2019}. For each selected water structure, one O-H bond is randomly picked and stretched with 20 different bond distances ranging from 0.74 {\AA} to 5.43 {\AA}. All the MRCI with Davidson correction using a relaxed reference (MRCI+Q) \cite{werner1988efficient, knowles1988efficient} energies for these datasets are generated using the default settings in \textsc{Molpro 2021.1}\cite{werner2020molpro} with a CASSCF reference space. For the \ch{H10} dataset, all the 49 structures and reference MRCI+Q-F12 energies are directly obtained from Ref.~\citenum{scherbela2022solving}. 

All the MOB features for the criegee, \ch{H10} chain, small radical, and water bond dissociation datasets are generated from \textsc{entos qcore} \cite{manby2019entos} software with the same feature generation protocol described in Husch et al.\cite{husch2020improved}. 
The density-fitted RHF/cc-pVTZ and RHF/cc-pVTZ-F12 \cite{kendall1992electron} computations are conducted for criegee and \ch{H10} datasets, respectively. The density-fitted ROHF computations for small radical and water bond dissociation datasets are conducted with cc-pVTZ and aug-cc-pVTZ \cite{kendall1992electron} basis sets, respectively. 
QM9\cite{ramakrishnan2014quantum,dipole_data} contains 133885 optimized structures of drug-like molecules with up to nine heavy atoms (HAs) of C, O, N, and F. QM7b-T and GDB-13-T \cite{Cheng2019,dataset,caltech_data} contain thermalized structures of 7211 molecules (7211 x 7) with at most seven HAs and 1000 molecules (1000 x 6) with thirteen HAs of C, O, N, Cl and S, respectively. The structures, reference data and features of QM9 and two thermalized organic molecule datasets are directly obtained from Ref.~\citenum{sun2022molecular} and Ref.~\citenum{husch2020improved}, respectively, and are the same as the ones used in the literature studies \cite{husch2020improved,sun2021molecular,cheng2022accurate,sun2022molecular}. For the QMSpin dataset, the singlet and triplet optimized structures and total MRCISD+Q-F12/cc-pVDZ-F12 energies are directly obtained from the QMSpin dataset. \cite{schwilk2020large,qmspin_data} The closed-shell features for singlet energies and the open-shell features for triplet energies are computed from RHF/cc-pVDZ and ROHF/cc-pVDZ using \textsc{entos qcore} \cite{manby2019entos}, respectively.

\subsection{MOB-ML model construction}
There is no need to perform feature selection to avoid overfitting since we adapt a kernel addition technique to construct a kernel directly, and all the diagonal MOB features are used in the training.
The training protocols in Eq.~\ref{eq:kernel_close} and Eq.~\ref{eq:kernel_open} are used for the closed-shell and open-shell KA-GPR models, respectively. The negative log marginal likelihood objective is minimized with respect to three kernel hyperparameters (one variance, one lengthscale, and one noise variance) using the L-BFGS optimizer with a convergence threshold of \num{1e-6}. All the training was performed on NVIDIA Tesla V100-SXM2-32GB GPUs with multi-GPU speed-ups. 
The example KA-GPR codes are available online at \url{https://github.com/
SUSYUSTC/KAGPR.git}.

\section{Results}
\subsection{MOB-ML for closed-shell systems with multi-reference characters: Criegee and \ch{H10} chain}

Figure~\ref{fig:criegee} first examines the performance of MOB-ML on the simplest criegee intermediate \ch{CH2OO}, a closed-shell molecule with multi-reference character with reference theories of CCSD(T)/cc-pVTZ and MRCI+Q/cc-pVTZ. The test set is fixed and contains 100 randomly selected geometries from the entire dataset, and the training sets include the randomly selected geometries from the 700 remaining structures. The predicted mean absolute errors (MAEs) are plotted versus the number of training structures. 
Since the average T1 and D1 diagnostics of the entire dataset are bigger than 0.02 (Mean=0.044) \cite{lee1995achieving,lee1995investigation,lee2003comparison} and 0.05 (Mean=0.179) \cite{langhoff1988ab}, respectively, the criegee molecule has multi-reference character and large non-dynamical correlation effects. Therefore, the results from single-reference theories, such as M{\o}ller-Plesset perturbation and coupled-cluster theories, are unreliable, and multi-reference treatments are required. \cite{lee1995achieving}
For CCSD(T), the results of directly learning total correlation energies by KA-GPR have similar accuracy to those of original MOB-ML pair-wise decomposition. This observation indicates that the KA-GPR has little or no accuracy loss to transfer between different configurations of the same molecule. 
For MRCI+Q, only the KA-GPR results are reported due to the non-trivial implementation of the MRCI pair energies. The chemical accuracy of 1 kcal/mol is achieved for both levels of theories with only 10 training geometries. By training on 700 geometries, accuracies of 0.0027 and 0.0045 kcal/mol are reached for CCSD(T) and MRCI+Q, respectively.

\begin{figure}[htbp]
    \centering
    \includegraphics[width=0.85\columnwidth]{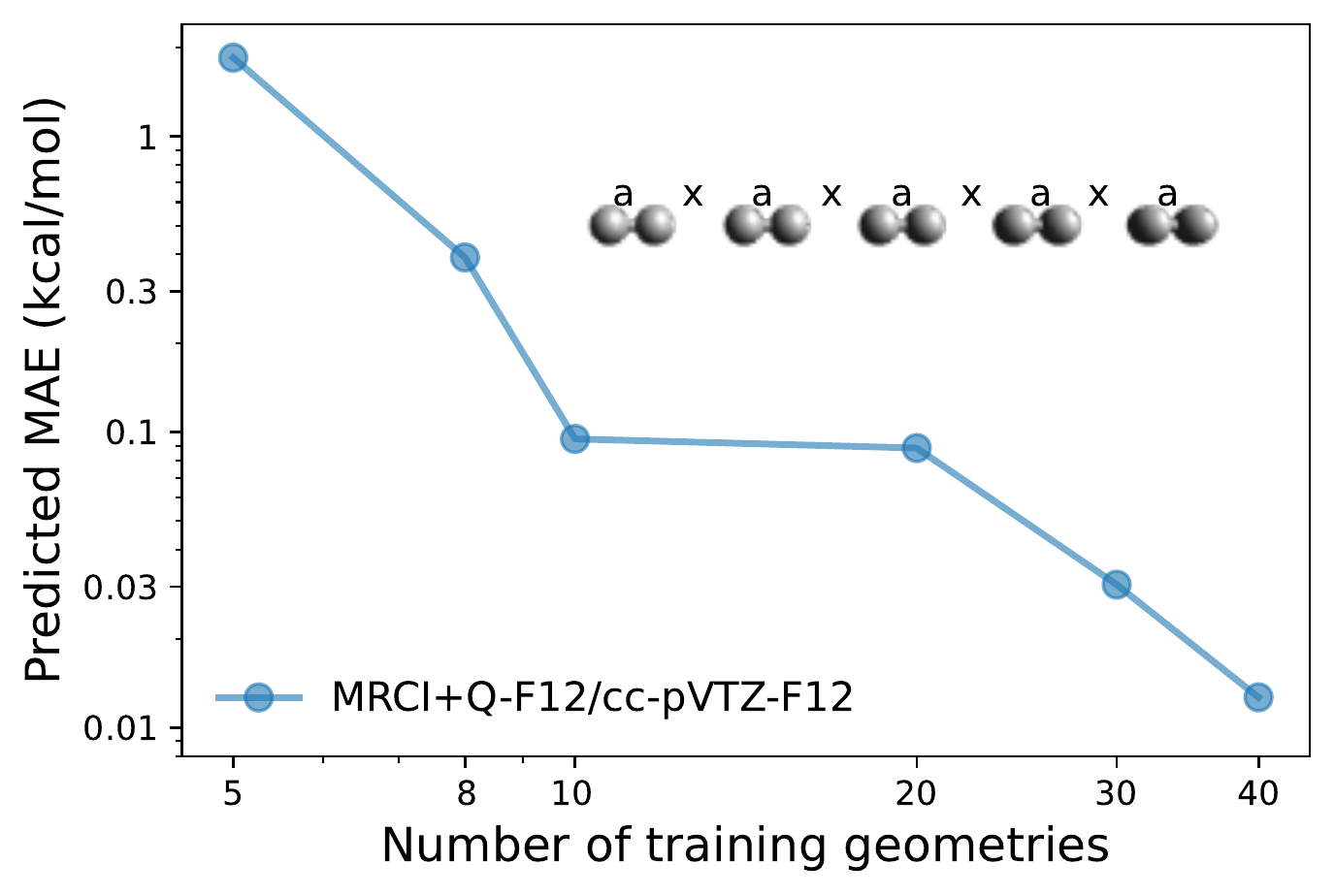}
    \caption{Mean absolute errors (MAEs) of predicted correlation energies for 9 test closed-shell \ch{H10} configurations using MOB-ML. An example structure of \ch{H10} are also shown, where $a$ is the bond length of each \ch{H2} and $x$ is the distance between two \ch{H2}.
    All the prediction MAEs are also listed in the Supporting Information Table S2.}    
    \label{fig:h10}
\end{figure}

We additionally test the performance of MOB-ML(KA-GPR) on the potential energy surface (PES) for the \ch{H10} chain in Fig.\ref{fig:h10}. The fixed test set contains 9 randomly selected structures, and the training sets are constructed as subsets from the reminder. The best MOB-ML(KA-GPR) model reaches 0.013 kcal/mol accuracy by training on 40 structures, which is significantly better than the MAE of DeepErwin (0.272 kcal/mol) on the same test geometries, though we note that DeepErwin uses a deep neural network-based ansatz for a variational Monte Carlo approach that works differently than MOB-ML. 

\subsection{Accurate predictions on benchmark closed-shell organic molecule datasets: QM9, QM7b-T, and GDB-13-T}
To further investigate the effects of changing learning protocol to KA-GPR on the accuracy and transferability of large closed-shell organic benchmark systems, including QM9, QM7b-T, and GDB-13-T, tested in the previous studies\cite{Cheng2019,cheng2019regression, husch2020improved,sun2021molecular,cheng2022accurate}. In this work, we compare the results of three MO-based or AO-based ML approaches using $Delta-$learning ideas, MOB-ML \cite{Welborn2018,Cheng2019,cheng2019regression,husch2020improved,sun2021molecular,cheng2022accurate,sun2022molecular}, OrbNet-Equi ($\Delta$-learning) \cite{orbnetequi} and QML (MO, $\Delta$-learning) \cite{Karandashev2022}, for these benchmark datasets. We note that QML (MO, $\Delta$-learning) also adapts the kernel addition technique in the kernel ridge regression to construct its models. 
\begin{figure}[htbp]
    \centering
    \includegraphics[width=0.85\columnwidth]{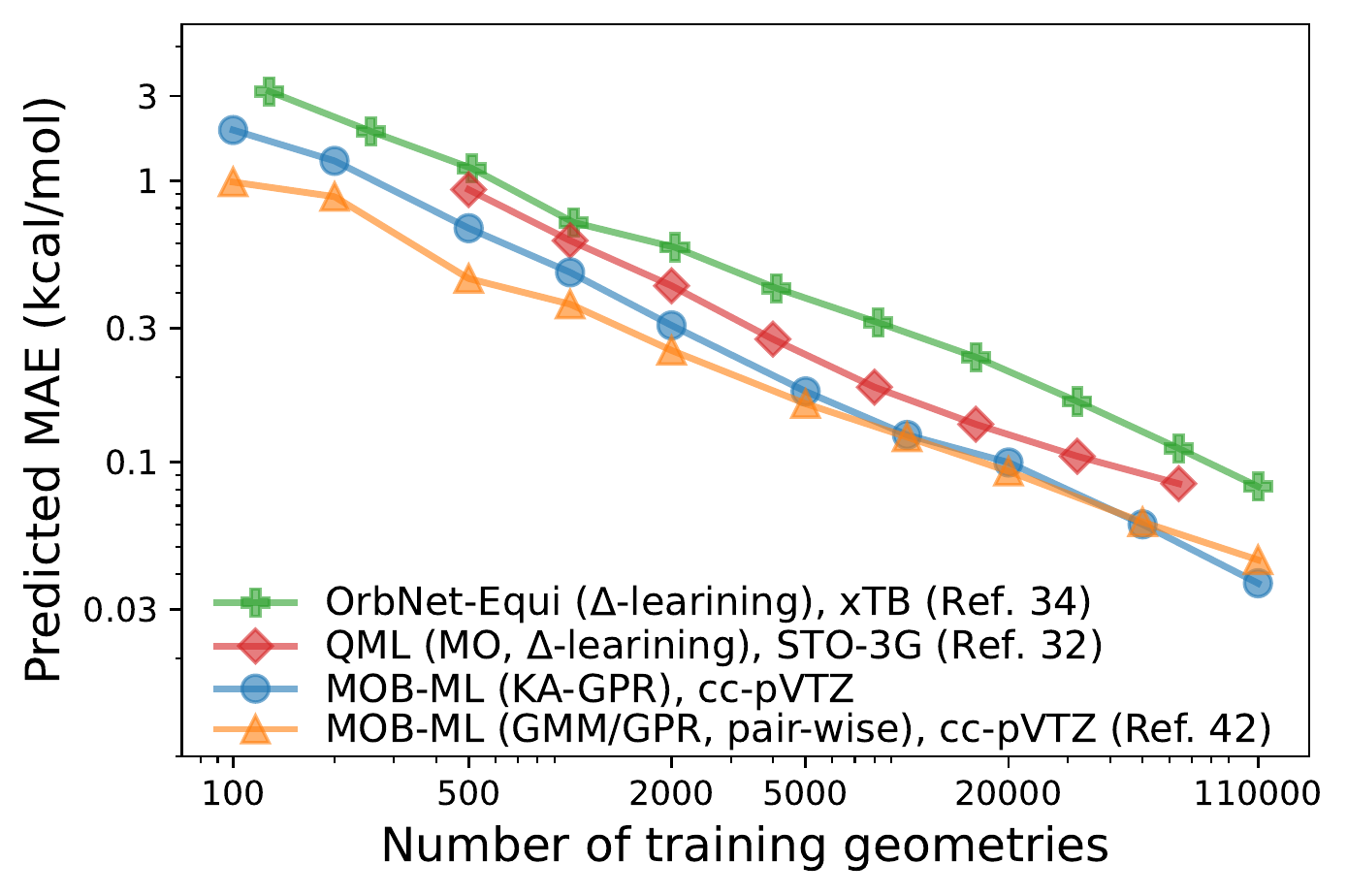}
    \caption{Learning curves of MOB-ML for test QM9 molecules at the MP2/cc-pVTZ level of theory. The models are generated with different numbers of randomly selected QM9 molecules. The learning curves of OrbNet-Equi \cite{orbnetequi} and QML \cite{Karandashev2022} using with a reference theory of B3LYP/6-31G(2df,p) \cite{ramakrishnan2014quantum} are also displayed for comparison. The results of MOB-ML (GMM/GPR) obtained from Ref.~\citenum{sun2022molecular} are also shown to compare with the ones from MOB-ML (KA-GPR). The corresponding the level of theory used for feature generation of each ML approach is also listed in the legend. The MAEs of MOB-ML (KA-GPR) are also listed in Table S3 in the Supporting Information.} 
    \label{fig:qm9}
\end{figure}

\begin{figure*}[bhtp]
    \centering
    \includegraphics[width=1.8\columnwidth]{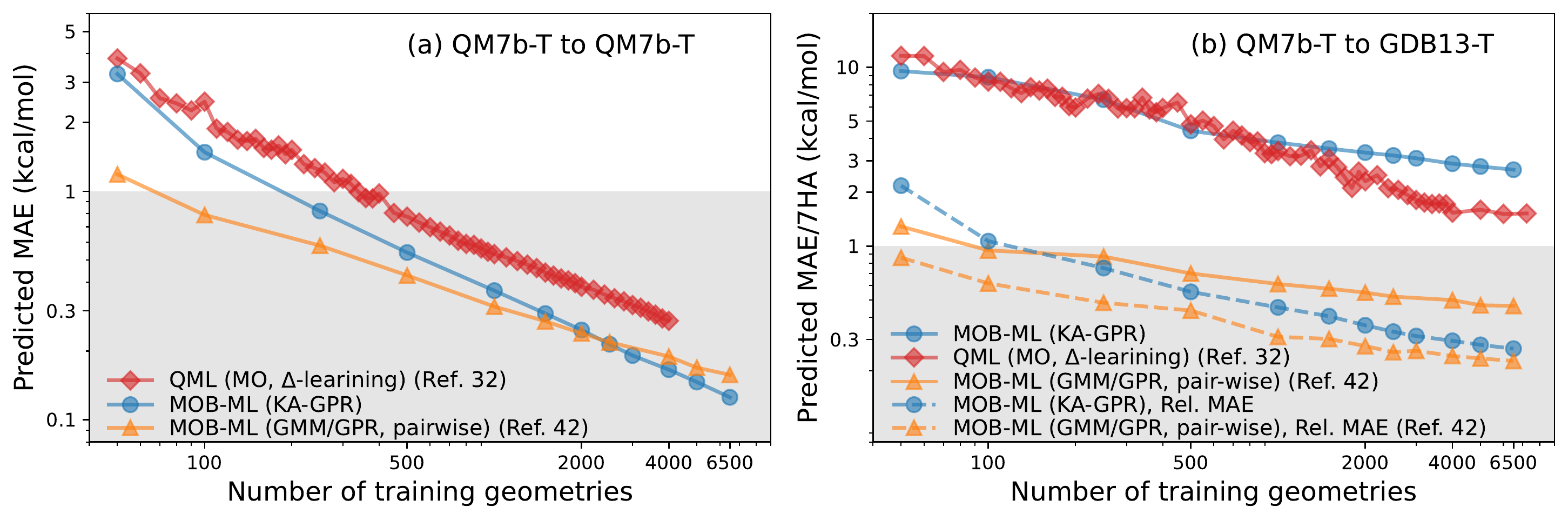}
    \caption{Learning curves for MP2/cc-pVTZ energy predictions from MOB-ML trained on QM7b-T and applied to (a) QM7b-T and (b) GDB-13-T. The literature results of learning pair energies by unsupervised clustering via Gaussian mixture and local regression with AltBBMM regressor in MOB-ML, i.e., MOB-ML (GMM/GPR, pair-wise), and QML (MO, $\Delta$-learning) using features generated from HF/cc-pVTZ computations are plotted. In panel (b), the prediction accuracies of the relative conformer energies (dash lines) computed by subtracting the corresponding true and predicted optimized structure energies of each GDB-13-T molecule are also shown.
    The corresponding computational cost of the feature generation in each ML approach is also listed in the legend, and the shaded areas correspond to the chemical accuracy region of 1 kcal/mol MAE.}    
    \label{fig:qm7b}
\end{figure*}
Figure~\ref{fig:qm9} plots the results of models trained and tested on the QM9 dataset using MOB-ML (KA-GPR) with a reference theory of MP2/cc-pVTZ. 
MOB-ML (GMM/GPR)\cite{sun2022molecular} and MOB-ML (KA-GPR) provide similar accuracies with all different training sizes, which verifies the conclusion that KA-GPR has little or no accuracy loss when transferring between different molecules with the same size. 
Both MOB-ML approaches perform better than OrbNet-Equi($\Delta$-learning) and QML (MO, $\Delta$-learning) across all the training sizes. This result might be attributed to the fact that the total correlation energies learnt by MOB-ML have a smaller variation than the energy differences between B3LYP/6-31G(2df,p) and HF/STO-3G or GFN-xTB. In addition, the features extracted from HF/cc-pVTZ computations might have a higher quality with more information content than the ones obtained from HF/STO-3G and GFN-xTB computations. 

To further assess the transferability of the KA-GPR learning protocol in MOB-ML, we train MOB-ML models on the QM7b-T dataset and test them on both QM7b-T and GDB-13-T datasets in Fig.~\ref{fig:qm7b} (a) and (b), respectively. In panel (a), the models are trained on the training sets containing only one conformer of the selected QM7b-T molecules and tested on the reminder conformers and molecules.  
For the training sizes smaller than 1000, MOB-ML models generated by learning pair energies provide significantly better accuracy than those generated by learning the total correlation energies. However, MOB-ML (KA-GPR) out-competes MOB-ML (GMM/GPR) when training over 3000 QM7b-T geometries. This is because the actual number of training points in the MOB-ML (KA-GPR) models is much fewer than the one in the MOB-ML (GMM/GPR) models. Both MOB-ML protocols have better accuracy than the state-of-the-art QML (MO, $\Delta$-learning) approach. The best MOB-ML (KA-GPR) model trained on 6500 QM7b-T geometries reaches an accuracy of 0.126 kcal/mol.

The same set of models trained on QM7b-T molecules are then applied to predict the GDB-13-T geometries in Fig.~\ref{fig:qm7b}(b) for all three ML approaches. MOB-ML (KA-GPR) has the same input features as MOB-ML (GMM/GPR) and a similar kernel computation protocol as QML (MO, $\Delta$-learning), and its MAEs are much worse than the ones of MOB-ML (GMM/GPR) but similar those of QML (MO, $\Delta$-learning). 
Learning the pair energies can ensure the accurate prediction of total correlation energies, as they are the sum of the pair energies. However, directly learning their sum can only guarantee to provide correct correlation energies but not accurate decomposed pair energies. 
Therefore, the pair-wise decomposition of the total correlation energies in MOB-ML is significant in ensuring the size extensivity and the transferability across different molecular sizes. 
However, both MOB-ML approaches are similarly accurate for the relative conformer energies computed by subtracting the optimized structure energies of each molecule, which suggests potential improvements of the KA-GPR protocol may be possible as a future research direction. The most accurate MOB-ML (KA-GPR) model provides an MAE/7HA = 0.266 kcal/mol for relative conformer energies of GDB-13-T, which is slightly worse than the value (MAE/7HA = 0.228 kcal/mol) from MOB-ML (GMM/GPR). 

\subsection{Performance of MOB-ML to learn energies of small organic radicals}

\begin{figure*}[htbp]
    \centering
    \includegraphics[width=1.9\columnwidth]{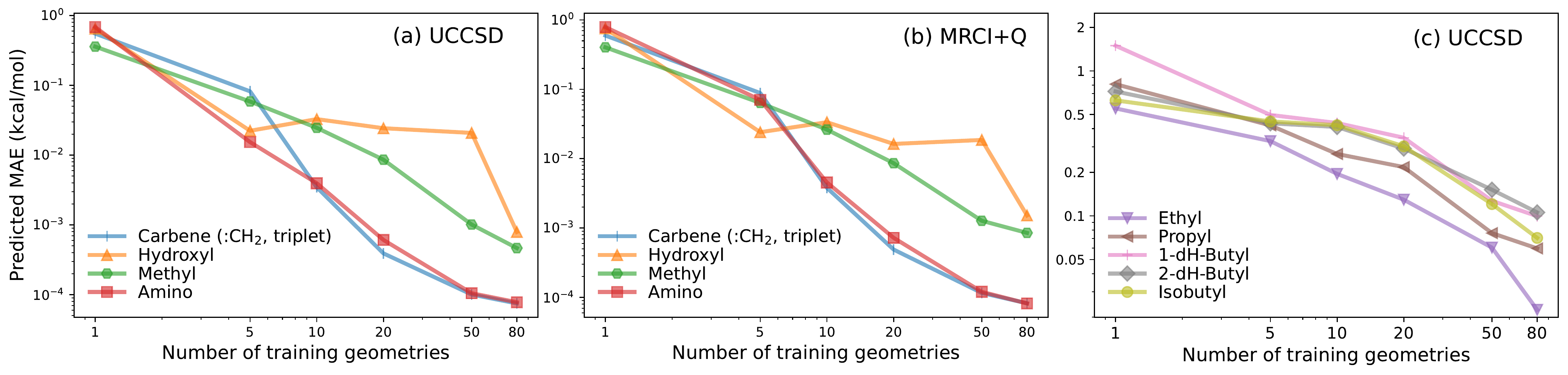}
    \caption{Mean absolute errors (MAEs) of predicted correlation energies for different test small radical structures using MOB-ML. The corresponding test set contains 120 test structures for each small radical molecule. Panel (a) and (c) display the results of KA-GPR at the level of LUCCSD/cc-pVTZ, and panel (b) shows the results at the level of MRCI+Q/cc-pVTZ. 
    For the reference theory of LUCCSD/cc-pVTZ, we display the results from learning the pair energies by GPR and total correlation energies by KA-GPR, respectively. 
    All the results are plotted on a logarithm scale.}    
    \label{fig:radical}
\end{figure*}

As the first set of examples for open-shell systems, we examine the performance of MOB-ML to learn the energies of nine different small radicals using the open-shell KA-GPR framework (Eq.~\ref{eq:kernel_open}). 
All the models are trained on randomly selected subsets with different sizes, and predict the corresponding fixed test sets with 100 geometries. 
MOB-ML (KA-GPR) reaches accuracies smaller than \num{1e-3} kcal/mol for all four radicals and both theories when training on 80 geometries in panel (a) and (b) of Fig~\ref{fig:radical}.
Panel (c) plots the results of five medium-size radicals with a reference theory of LUCCSD/cc-pVTZ, and MOB-ML (KA-GPR) performs reasonably well for all of these medium-size radicals. Except for the 1-dH-butyl radical, eight other tested radicals reached the chemical accuracy of 1 kcal/mol with only one training example.

\subsection{MOB-ML for water bond dissociation PES}
We then extend the KA-GPR framework to a more challenging system, the OH bond dissociation PES of water, computed at the MRCI+Q/aug-cc-pVTZ level of theory. We selected 500 structures from the water bond dissociation dataset as the test set. In this test set, 480 of the 500 are randomly selected from the subset generated 49 initial AIMD structures with 20 different bond lengths (980 in total), and the remaining 20 structures are on one test water molecule bond dissociation PES. 

In MOB-ML, we use ROHF with a spin multiplicity=3 instead of RHF with spin multiplicity=1 as the reference to predict the singlet MRCI+Q ground state.
RHF does not provide meaningful MOB features due to the qualitatively wrong descriptions of the electronic structure when the O and H are distant enough.
Meanwhile, ROHF could always generate correct descriptions for triplet states over the entire PES, which also qualitatively describe the molecular information of each geometry.
Since the exact singlet and triplet state energies coincide at an infinite O-H distance using ROHF reference in MOB-ML is expected to give better predictions even for the singlet ground state energies. The corresponding dissociation curves are shown in Fig. S1 of Supporting Information. In principle, the MOB-ML model could be learnt by two sets of features for bond dissociation: 
\begin{equation}
    \label{eq:app}
    \begin{aligned}
    E_\text{corr}^\text{ML} &\approx
        e^\text{RHF} \left[\ff_{ij, \text{RHF}}\right]
    \\
    E_\text{corr}^\text{ML} &\approx
        e^\text{ROHF} \left[\ff_{ij, \text{ROHF}}\right]
    \end{aligned}
\end{equation}
The definitions of correlation energies here are extended  to the differences between the total energy and RHF or ROHF energies.
Since it is more challenging to get the dissociated region correct, we use the open-shell features generated by ROHF calculations and predict the energy differences between ROHF and true MRCI+Q energies with correct reference wave functions.

\begin{figure}[htbp]
    \centering
    \includegraphics[width=0.9\columnwidth]{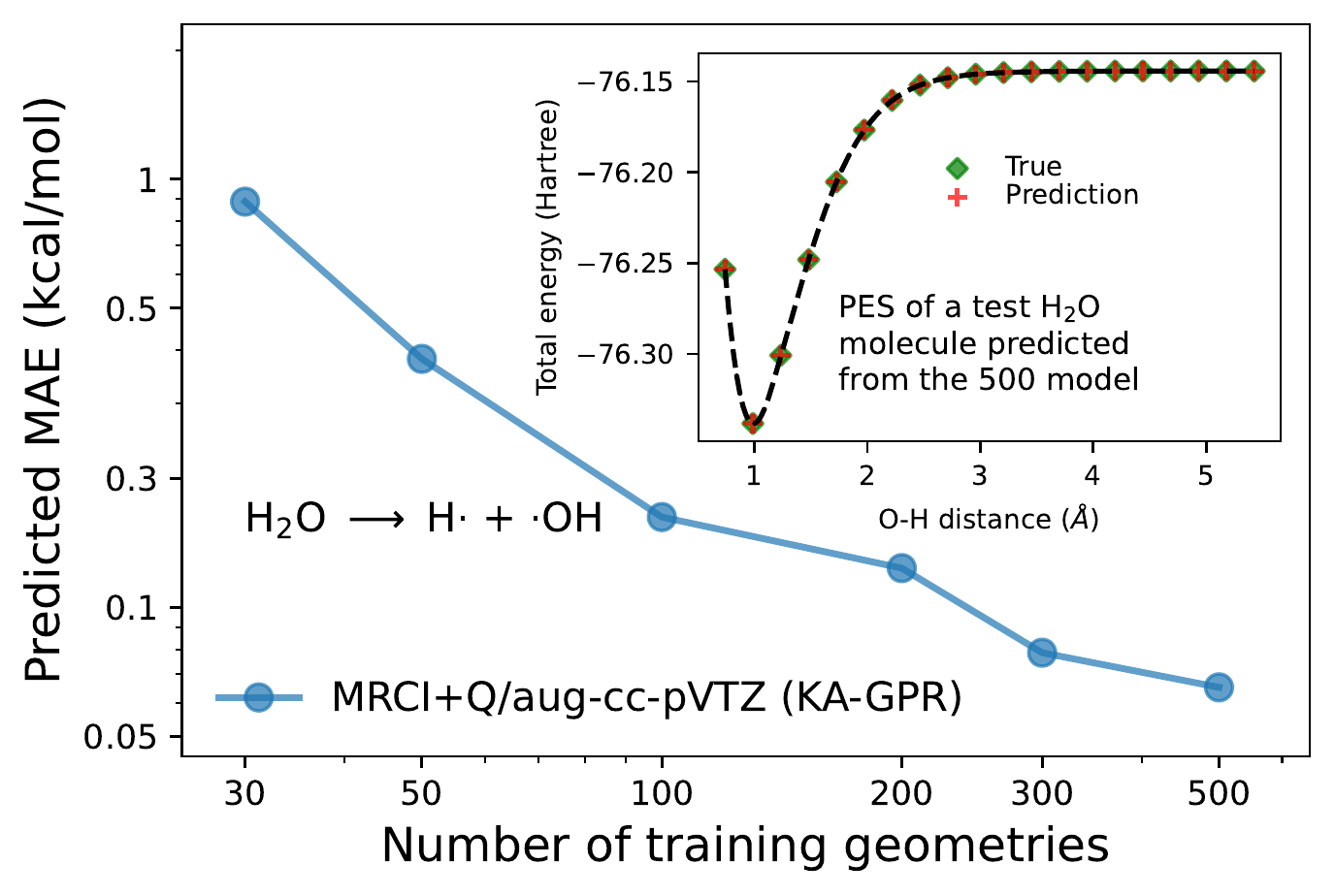}
    \caption{Learning curve of MOB-ML on thermalized water bond breaking datasets with a reference theory of MRCI+Q/aug-cc-pVTZ. The features are computed using ROHF/aug-cc-pVTZ. The true and predicted bond dissociation curves from the best MOB-ML model for a test thermalized water structure is also shown in the figure. We note that none of the points on this dissociation curve is included in the training set.}    
    \label{fig:water_bond}
\end{figure}

Figure ~\ref{fig:water_bond} plots the learning curve of the single OH bond dissociation dataset and the bond dissociation PES of one test water structure. The chemical accuracy of 1 kcal/mol is reached by training on 30 geometries. The most accurate model learnt on 500 training geometries provides an MAE of 0.065 kcal/mol. In addition, the predicted PES from the best model of the test water molecule is nearly identical to the true PES.

\subsection{Performances of MOB-ML on the QMSpin dataset: Large carbenes derived from QM9 molecules}

We finally apply MOB-ML (KA-GPR) to learn the energies and spin gaps in QMSpin dataset\cite{schwilk2020large,qmspin_data}, which includes 5022 optimized singlet and 7958 optimized triplet carbene structures generated from QM9 dataset. All the structures are optimized with open-shell restricted B3LYP/def2-TZVP, and most of these structures have computed singlet and triplet energies at MRCISD+Q-F12/
cc-pVDZ-F12, resulting in around 26000 energy points in total. For the test set, we randomly select 1000 test molecules that have singlet and triplet energies computed for both singlet and triplet structures (2000 structures and 4000 energies in total). The vertical and adiabatic spin gap energies could be computed by
\begin{align}
    E_{\text{singlet}}^\text{gap}=E_{\text{singlet}}^t-E_{\text{singlet}}^s,\\
    E_{\text{triplet}}^\text{gap}=E_{\text{triplet}}^t-E_{\text{triplet}}^s,\\
    E_{\text{adiabatic}}^\text{gap}=E_{\text{triplet}}^t-E_{\text{singlet}}^s,
\end{align}
where $E_{\text{singlet}}^s, E_{\text{singlet}}^t, E_{\text{triplet}}^s, E_{\text{triplet}}^t$ are the singlet energies of singlet structures, the triplet energies of singlet structures, the singlet energies of triplet structures, the triplet energies of triplet structures, respectively.
The singlet and triplet energy models are learnt on the same sets of training QMSpin structures. 

\begin{figure*}[htbp]
    \centering
    \includegraphics[width=1.85\columnwidth]{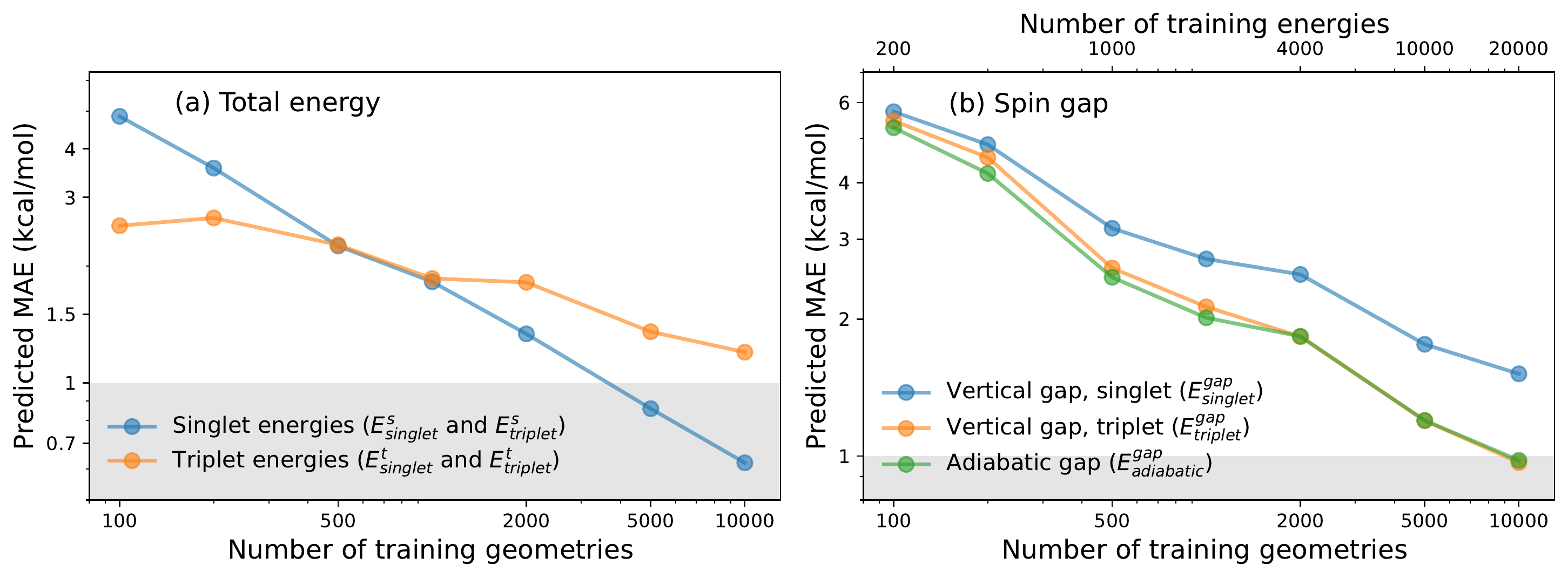}
    \caption{Performances of MOB-ML on (a) singlet and triplet total energies and (b) spin gap energies of QMSpin structures. The test set contains 1000 singlet and 1000 triplet optimized structures (2000 in total) that are generated from the same 1000 original QM9 structures. These test singlet and triplet optimized structures have the energies computed with both singlet and triplet multiplicities, therefore can directly compute the vertical and adiabatic gap energies. The singlet and triplet energy models are learnt separately by MOB-ML (KA-GPR) in panel (a), and the predicted vertical gaps for singlet and triplet structures and adiabatic gaps are computed using these two sets of models are plotted in panel (b). Both the number of training geometries and energies are shown in panel (b).
    All the MAEs are plotted as functions of the number of training geometries or energies on a logarithm scale, and the shaded areas correspond to the chemical accuracy region of 1 kcal/mol MAE.}    
    \label{fig:qmspin}
\end{figure*}

Figure ~\ref{fig:qmspin} displays the prediction accuracies of singlet and triplet energy models and the corresponding spin gap energies in panels (a) and (b), respectively. The singlet energy model has a steeper learning curve than the triplet energy model, suggesting that learning open-shell molecules is more complicated than learning closed-shell molecules. The best accuracies of MOB-ML (KA-GPR) by training on 10000 QMSpin structures are 0.623 kcal/mol and 1.120 kcal/mol for the singlet energy and triplet energy models, respectively. 
There is only one literature ML approach, SpookyNet\cite{unke2021spookynet}, testing on this QMSpin dataset. 
By training on 20000 energy points in total, the MAE of MOB-ML (KA-GPR) for the entire test dataset is 0.872 kcal/mol, which outcompetes the result from SpookyNet with electronic embeddings (1.57 kcal/mol).
In panel (b), the prediction accuracies of MOB-ML (KA-GPR) for the vertical spin gap of triplet structures and the adiabatic gap reach the chemical accuracy of 1 kcal/mol. The vertical gap of singlet structures reaches an MAE of 1.517 kcal/mol, which is slightly worse than the results of two other gaps.

\section{Conclusion}
In this work, we introduce a novel training protocol in MOB-ML, KA-GPR, to extend MOB-ML to learning the correlation energies of closed-shell and open-shell molecular systems computed with general electronic structure theories (using MRCI as an example) by directly training the total correlation energy without pair-wise decomposition.
The performances of MOB-ML (KA-GPR) are demonstrated on various chemical systems, including a closed-shell molecule with multi-reference characters (criegee), a PES of a strongly-correlated system (\ch{H10} chain), large organic benchmark datasets (QM9, QM7b-T, and GDB-13-T), nine small open-shell radicals, water bond dissociation curves, and large carbene datasets (QMSpin). MOB-ML (KA-GPR) provides state-of-the-art prediction accuracies for all the tested applications. Although the predicted relative conformer energies remain accurate, the usage of KA-GRP leads to a significant loss in transferability for the absolute total energies when applying the models learnt on QM7b-T molecules to predict the GDB-13-T molecules. This work enables the possibility of MOB-ML to learn molecular energies with accuracy beyond CCSD(T) level. In addition, the near-exact predicted single bond dissociation curve for a test water molecule and accurately predicted large carbene energies in QMSpin using MOB-ML (KA-GPR) also indicate a promising application of MOB-ML as a PES to study challenging organic chemistry reactions involving radicals. 

\begin{acknowledgments}
We thank Tamara Husch for helpful discussions at the early stages of this project. TFM acknowledges support from the US Army Research Laboratory (W911NF-12-2-0023), the US Department of Energy (DE-SC0019390), the Caltech DeLogi Fund, and the Camille and Henry Dreyfus Foundation (Award ML-20-196). Computational resources were provided by the National Energy Research Scientific Computing Center (NERSC), a DOE Office of Science User Facility supported by the DOE Office of Science under contract DE-AC02-05CH11231.
\end{acknowledgments}

\section*{Supporting Information}
The structures, HF energies, and reference energies and additive MOB features of criegee, small radicals, water bond dissociation and QMSpin datasets are available at Caltech Data\cite{mrci_data}: \url{https://data.caltech.edu/records/20200}. The features and energies of QM9 and thermalized organic molecule datasets (QM7b-T and GDB-13-T) are directly obtained from Ref.~\citenum{sun2022molecular} and Ref.~\citenum{husch2020improved}. The implementation of the multi-GPU AltBBMM and GMM are available online at \url{https://github.com/SUSYUSTC/
KAGPR.git}. All the data plotted in the figures are also listed in the tables in the Supporting Information.

\bibliography{main}
\end{document}